\documentclass[prl,amsmath,amssymb,showpacs,superscriptaddress,nofootinbib,twocolumn]{revtex4-1}
\usepackage{graphicx}
\usepackage{epstopdf}
\usepackage[mathlines]{lineno}
\usepackage{dcolumn}
\usepackage{color}
\usepackage{bm}
\usepackage{overpic}
\usepackage{rotating}
\usepackage{times}
\usepackage{multirow}
\usepackage{float}
\usepackage{mathrsfs}
\usepackage{booktabs}
\usepackage[mathlines]{lineno}
\newcommand{\pp}{\pi^+\pi^-}

\newcommand{\LL}{\ell^+ \ell^-}
\newcommand{\EE}{e^+e^-}

\newcommand{\MM}{\mu^+\mu^-}

\newcommand{\psip}{\psi^{\prime}}

\newcommand{\jpsi}{J/\psi}

\newcommand{\X}{X(3872)}
\newcommand{\DDstar}{D^{*0}\bar{D^{0}}}
\newcommand{\Dstar}{D^{*0}}
\newcommand{\DO}{D^{0}}
\newcommand{\DDbar}{D^{0}\bar{D^{0}}}
\newcommand{\B}{\mathcal{B}}
\newcommand{\DpDm}{D^{+}D^{-}}

\newcommand{\gammaH}{\gamma_{\rm H}}
\newcommand{\gammaL}{\gamma_{\rm L}}

\newcommand{\chisq}{\chi^{2}}

\newcommand{\Rpsi}{R_{\gamma\psi}}
\newcommand{\psipOVERjpsi}{\frac{\B(\X\to\gamma\psip)}{\B(\X\to\gamma\jpsi)}}
\begin{document}
\hyphenpenalty=10000
\tolerance=1000
\normalsize
\parskip=5pt plus 1pt minus 1pt

\title{\boldmath Study of open-charm decays and radiative transitions of the X(3872)}
\author{
M.~Ablikim$^{1}$, M.~N.~Achasov$^{10,e}$, P.~Adlarson$^{63}$, S. ~Ahmed$^{15}$, M.~Albrecht$^{4}$, A.~Amoroso$^{62A,62C}$, Q.~An$^{59,47}$, ~Anita$^{21}$, Y.~Bai$^{46}$, O.~Bakina$^{28}$
% ...
\\
\vspace{0.2cm}
(BESIII Collaboration)\\
\vspace{0.2cm} {\it
$^{1}$ Institute of High Energy Physics, Beijing 100049, People's Republic of China\\
% ...
$^{4}$ Bochum Ruhr-University, D-44780 Bochum, Germany\\
% ...
$^{10}$ G.I. Budker Institute of Nuclear Physics SB RAS (BINP), Novosibirsk 630090, Russia\\
% ...
$^{15}$ Helmholtz Institute Mainz, Johann-Joachim-Becher-Weg 45, D-55099 Mainz, Germany\\
% ...
$^{21}$ Indian Institute of Technology Madras, Chennai 600036, India\\
% ...
$^{28}$ Joint Institute for Nuclear Research, 141980 Dubna, Moscow region, Russia\\
% ...
$^{46}$ Southeast University, Nanjing 211100, People's Republic of China\\
$^{47}$ State Key Laboratory of Particle Detection and Electronics, Beijing 100049, Hefei 230026, People's Republic of China\\
% ...
$^{59}$ University of Science and Technology of China, Hefei 230026, People's Republic of China\\
% ...
$^{62}$ (A)University of Turin, I-10125, Turin, Italy; (B)University of Eastern Piedmont, I-15121, Alessandria, Italy; (C)INFN, I-10125, Turin, Italy\\
$^{63}$ Uppsala University, Box 516, SE-75120 Uppsala, Sweden\\
% ...
$^{e}$ Also at the Novosibirsk State University, Novosibirsk, 630090, Russia\\
% ...
}
}
\date{\today}

\begin{abstract}

  The processes $X(3872)\to D^{*0}\bar{D^{0}}+c.c.,~\gamma J/\psi,~\gamma \psi(2S),$ and $\gamma D^{+}D^{-}$ are searched for in
  a $9.0~\rm fb^{-1}$ data sample collected at center-of-mass energies between $4.178$ and $4.278$~GeV with the BESIII detector.
  We observe $X(3872)\to D^{*0}\bar{D^{0}}+c.c.$ and find evidence for $\X\to\gamma J/\psi$ with statistical significances of $7.4\sigma$
  and $3.5\sigma$, respectively. No evident signals for $X(3872)\to\gamma\psi(2S)$ and $\gamma D^{+}D^{-}$ are found, and upper limit on the
 relative branching ratio $R_{\gamma \psi} \equiv\frac{\mathcal{B}(X(3872)\to\gamma\psi(2S))}{\mathcal{B}(X(3872)\to\gamma J/\psi)}<0.59$
  is set at 90\% confidence level.
  Measurements of branching ratios relative to decay $X(3872)\to\pi^+\pi^- J/\psi$ are also reported for decays $X(3872)\to D^{*0}\bar{D^{0}}+c.c., ~\gamma\psi(2S),~\gamma J/\psi$, $\gamma D^{+}D^{-}$, as well as the non-$\DDstar$ three-body decays
  $\pi^0\DDbar$ and $\gamma\DDbar$.

\end{abstract}

\maketitle


Since the discovery of the $\X$ in 2003~\cite{FirstObservationOfX} by the Belle Collaboration, many properties of this exotic state have been reported,
including its mass, an upper limit (UL) on its width, and its $J^{PC}$ quantum numbers~\cite{JPC_X3872_1,JPC_X3872_2}.
The ratio of the branching fraction (BF) of $\X\to\gamma\psi'$ (in this paper we use the notation $\psip$ to denote the $\psi(2S)$ resonance) to
$\X\to\gamma\jpsi$, $\Rpsi\equiv\psipOVERjpsi$, is predicted to be
in the range $(3\sim4)\times10^{-3}$ if the $\X$ is a $\DDstar$ molecule~\cite{DDstarMolecule,R_in_DDstar},
$0.5\sim5$ if it is a molecule-charmonium mixture~\cite{R_in_mix},
and
$1.2\sim15$ if it is a pure charmonium state~\cite{R_in_chm1,R_in_chm2,R_in_chm3,R_in_chm4,R_in_chm5,R_in_chm6,R_in_chm7}.
LHCb reported a $4.4\sigma$ evidence for the decay $\X\to\gamma\psip$ with
$\Rpsi=2.46\pm0.64\pm0.29$~\cite{R_in_lhcb},
which is in good agreement with the BaBar result $R_{\gamma\psi}=3.4\pm1.4$~\cite{R_in_babar}.
On the other hand, the Belle Collaboration report an upper limit of $\Rpsi<2.1$ at the
90\% confidence level (C.L.)~\cite{R_in_belle}.
$\X$ is produced at BESIII via the radiative decay from the $Y(4260)$ state~\cite{X2pipiJpsi_BESIII,BESIII_omegaJpsi} with a background level lower than at other experiments.
This makes BESIII particularly well suited for studies of $\X$ decays to final states containing photons and $\pi^0$ mesons.

With BESIII we cannot measure absolute BFs of $\X$ decays since the cross section of $\EE\to\gamma\X$ is unknown.
Instead we determine their ratios to the $\pp\jpsi$ mode.
As discussed in Ref.~\cite{DDstarMolecule}, the BF ratio of $\frac{\B(\X\to\DDstar+c.c.)}{\B(\X\to\pp\jpsi)}$
can be reliably calculated if the $\X$ is a weakly-bound molecule, in which case the ratio is predicted to be around
$0.08$ for a binding energy of 0.7~MeV.
Additionally, the decay width to $\gamma\DpDm$ is predicted to be $0.2~\rm keV$ for the molecular case.

In this paper, we report the study of $\X\to\DDstar$, $\gamma\jpsi$, $\gamma\psip$, and $\gamma\DpDm$ using $\EE$ annihilation data collected with the BESIII detector at center-of-mass energies ranging from $4.178$ to $4.278$ GeV.
The total integrated luminosity is $9.0~\rm fb^{-1}$. Charge-conjugate modes are implied throughout.
A detailed description of the BESIII detector and the upgrade of the time-of-flight system can be found in
Refs.~\cite{Ablikim:2009aa,etof}.

Monte Carlo (MC) simulated event samples are produced with a {\sc geant4}~\cite{geant4} based framework.
Large simulated samples of generic $\EE\to$hadrons events, which in total are 40 times the size of the data sample, are used to estimate background conditions.
The simulation of inclusive MC samples is described in Ref.~\cite{gmc}.
The signal process $\EE\to\gamma\X$ is generated assuming it is a pure electric dipole (E1) transition, and
the subsequent $\X$ decays are generated uniformly in the phase space except $\X\to\gamma\jpsi~(\gamma\psip)$ which is
generated assuming a pure E1 transition too.
The $\X$ resonance is described with a Flatt\'{e} formula with  parameter values taken from Ref.~\cite{Zhang:2009bv}.

When selecting $\X\to\gamma\jpsi$ decays, we use lepton pairs ($\LL,~\ell=e,~\mu$) to reconstruct the $\jpsi$,
while for the $\X\to\gamma\psip$ selection, we exploit the decays $\psip\to\pp\jpsi~(\jpsi\to\LL)$ and $\psip\to\MM$.
We use the same selection criteria for the charged
tracks and photons as described in Ref.~\cite{BESIII_omegaJpsi}.
The invariant mass of the lepton pair is required to be $|M(\LL)- m_{\jpsi(\psip)}|<0.02~\rm GeV/$$c^2$
for the $\jpsi$ or $\psip$ selection.
We use throughout this paper the notation $m_{\rm particle}$ to represent the mass of the specific particle listed in the PDG~\cite{PDG}.
In the case of $\X$ decays to charmed mesons, the $D^{*0}\to\gamma D^{0}$ and $\pi^0 D^{0}$ decays
are used to reconstruct the $D^{*0}$.
The $D^{0}$ is reconstructed via its
$K^-\pi^+,~K^-\pi^+\pi^0$, and $K^-\pi^+\pi^+\pi^-$ decay modes,
while the $D^{+}$ is reconstructed via its $K^-\pi^+\pi^+$ and $K^-\pi^+\pi^+\pi^0$ modes.
The particle identification (PID) of kaons and pions is based on the $dE/dx$ and time of flight information.
Assumption of a given particle identification is based on the larger of the two PID hypotheses probabilities.

A kinematic fit is performed to the event, with the constraints on the masses of the $\pi^{0}$, $D^{\pm/0}$ candidates, and the initial four momentum of the colliding beams.
When there are ambiguities due to multi-photon candidates in the same event,
we choose the combination with the smallest $\chi^2$ from the kinematic fit.
The $\chisq$ of the kinematic fit is required to be less than 40 for $\X\to\gamma\jpsi$, and less than 60 for the other modes.
In addition, the $\chisq$ of the kinematic fit of the hypothesis under study should be smaller than those for hypotheses with extra or fewer photons.
For all channels other than $\X\to\pi^0\DDbar$, there are two radiative photons. One is produced in $\EE$ annihilation directly and the other from $X(3872)$ or $D^*$ decay.
We denote the photon with larger energy after the kinematic fit as $\gamma_{\rm H}$ and the other $\gamma_{\rm L}$.
In these decays, $\pi^0$ and $\eta$ vetoes are imposed on the invariant mass of the photon pair $M(\gammaL\gammaH)$ to suppress further the possible $\pi^0$ and $\eta$ background,
$i.e.$, $|M(\gammaL\gammaH)-m_{\pi^0(\eta)}|>0.02(0.03)~\rm GeV/$$c^2$.

For the decay $\X\to\gamma\jpsi$, studies performed on  the inclusive MC sample indicate that the dominant backgrounds are Bhabha and di-muon events for $\jpsi\to\EE$ and $\MM$, respectively.
To suppress Bhabha events in the  $\jpsi\to\EE$ selection, the cosine of the polar angle of the selected photons, $\cos\theta$, is required to be within the interval $[-0.7, 0.7]$.
For $\sqrt{s}$=4.178---4.278 GeV, the energy of the photon from $\EE\to \gamma X(3872)$ is always lower than that from $X(3872)\to \gamma J/\psi$.
Background from $\EE\to\gamma\chi_{c1,2}$ with $\chi_{c1,2}\to\gamma\jpsi$
is suppressed by requiring $\left|M(\gammaL\jpsi)-m_{\chi_{c1,2}}\right|>0.02~\rm GeV/$$c^2$.
Here and below, $M(\gamma_{H/L}\jpsi)\equiv M(\gamma_{H/L}\LL)-M(\LL)+m_{\jpsi}$.
Neither peaking nor $\chi_{c1,2}$ background is found in the $M(\gamma_{H}\jpsi)$ spectra.

\begin{figure}[htbp]
\raggedright
    \begin{overpic}[width=0.22\textwidth]{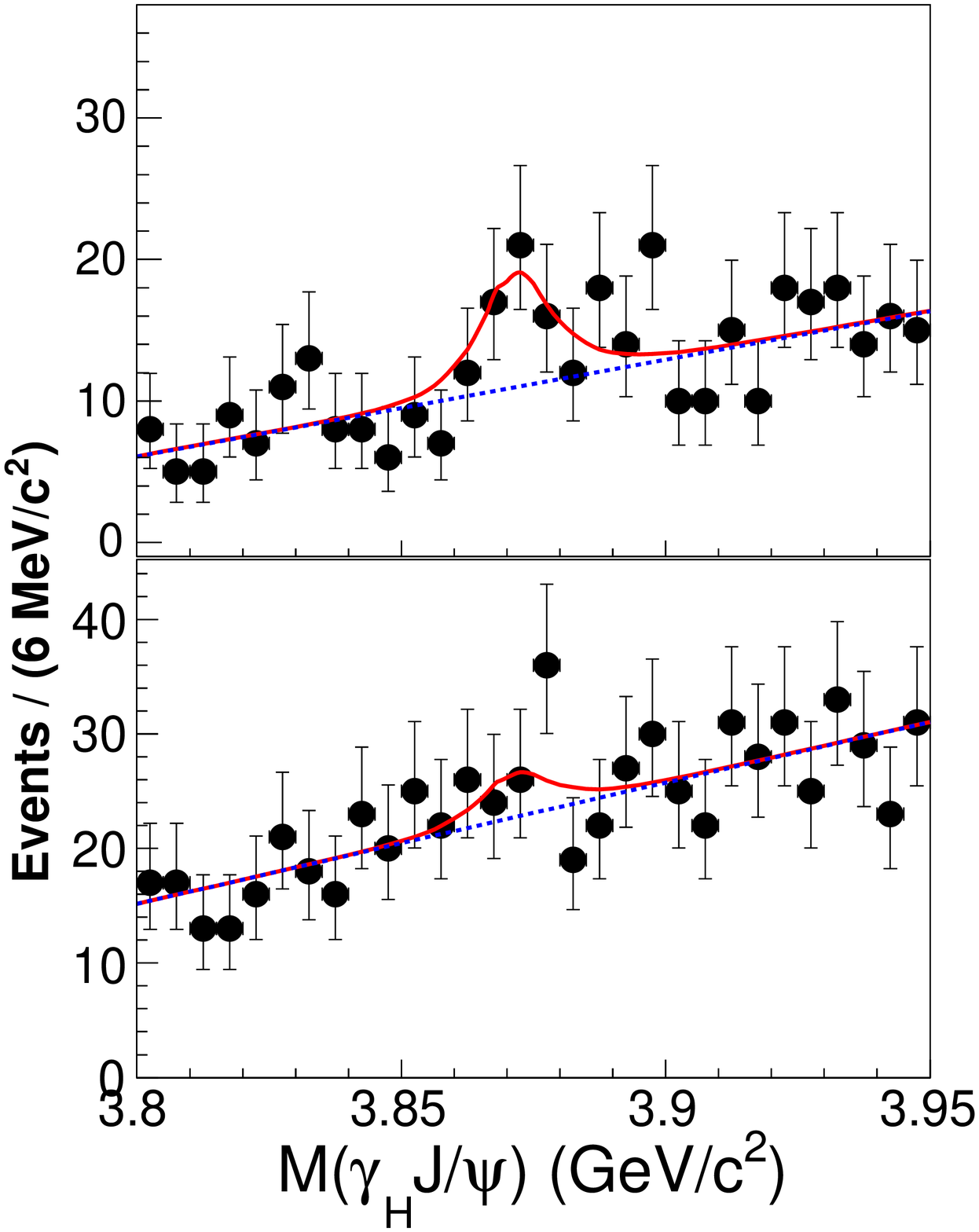}
        \put(20,85){\textbf{(a)}}
    \end{overpic}
    \begin{overpic}[width=0.22\textwidth]{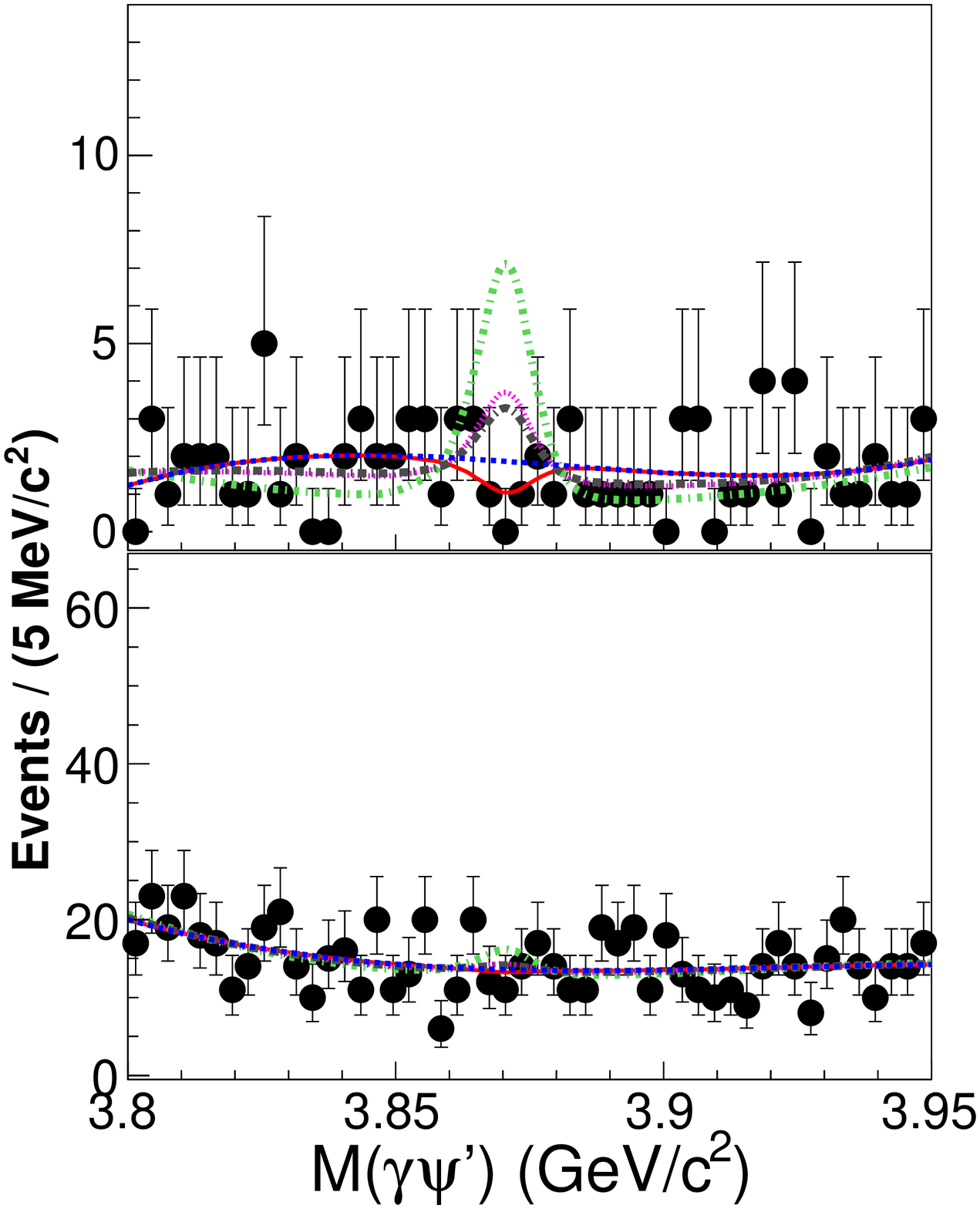}
        \put(20,85){\textbf{(b)}}
    \end{overpic}
\caption{(a) Fit results for $\X\to\gamma\jpsi$ for the $\MM$ (top) and $\EE$ (bottom) mode. (b)  Fit results for $\X\to\gamma\psip$ for the $\pp\jpsi$ (top) and $\MM$ (bottom) mode. The points with error bars are from data, the red curves are the best fit (color online). In (b), the rose-red dotted line represents the fit with the signal constrained to the expectation using $\X\to\pp\jpsi$ based on the relative ratios taken from a global fit~\cite{Li:2019kpj}; the green dash-dotted lines are using $\X\to\gamma\jpsi$ as the reference based on the LHCb measurement~\cite{R_in_lhcb}, and the grey long dashed lines are using $\X\to\gamma\jpsi$ as the reference based on the Belle measurement~\cite{R_in_belle}.}\label{fig:FitX}
\end{figure}

To obtain the number of signal events, a simultaneous fit is performed on the mass spectra of $\gammaH\jpsi$ with $\jpsi\to\MM$ and $\EE$.
Throughout this paper, we use an unbinned maximum-likelihood fit as the nominal fit method.
The ratio of signal yields for $\MM$ and $\EE$ modes is constrained to the ratio of the corresponding BFs, corrected by the ratio of the corresponding reconstruction efficiencies.
In the fit, the signal distributions are described with shapes obtained from the MC simulation, and the backgrounds are described with a second-order Chebychev polynomial.
The signal yield, background normalization, and coefficients of the polynomial are free in this fit and the other fits in this paper.
The distributions of $M(\gammaH\jpsi)$ as well as the fit results are shown in Fig.~\ref{fig:FitX}(a).
The statistical significance for $\X\to\gamma\jpsi$ is always greater than $3.5\sigma$, evaluated with a range of alternative background shapes.
The significance is calculated by comparing the likelihoods with and without the signal components included, and taking the change in the number of degrees of freedom (ndf) into account.
From the fit we obtain $(20.1\pm6.2)\times10^{2}$ BF- and efficiency-corrected $\X\to\gamma\jpsi$ events, corresponding to
$38.8\pm11.9$ and $18.4\pm5.6$ events for $\jpsi\to\MM$ and $\EE$, respectively.
The goodness of the fit is $\chi^2/{\rm ndf}=27.8/52~(p=1.0)$.

For the decay $\X\to\gamma\psip$ with $\psip\to\MM$, the selection criteria for $\psip\to\MM$ are analogous as those for $\jpsi\to\MM$.
For the $\psip\to\pp\jpsi$ channel, we select events with the corrected mass $M(\pp\jpsi)\equiv M(\pp\LL)-M(\LL)+m_{\jpsi}$ satisfying
$|M(\pp\jpsi) - m_{\psip}|<0.006~\rm GeV/$$c^2$ as the signal-event candidates.
The main background is $\EE\to\pp\psip$, with $\psip\to\gamma\gamma\jpsi$.
We require $|M(\pp)_{\rm recoil} - m_{\psip}| > 0.01~\rm GeV/$$c^2$ to suppress these events, where $M(\pp)_{\rm recoil}$ is the recoiling mass of the $\pp$ system.

To determine the number of $\X\to\gamma\psip$ decays, similar fits are performed to the invariant mass $M(\gamma\psip)$ distribution as described above, where $\gamma$ includes $\gammaL$ and $\gammaH$.
The distribution of $M(\gamma\psip)$ as well as the fitting results are shown in Fig.~\ref{fig:FitX}(b).
The fit yields $(-1.1\pm5.2)\times10^2$ BF- and efficiency-corrected $\X\to\gamma\psip$ events,  corresponding to
$-0.9\pm4.1$ and $-0.4\pm1.6$ $\psip\to\pp\jpsi$ and $\MM$ events, respectively,
and the goodness of the fit is $\chi^2/\rm ndf=45.0/58~({\it p}=0.89)$.
The UL of the number of BF- and efficiency-corrected events is calculated to be $1.0\times10^3$ at the 90\% C.L..
This is obtained by integrating the likelihood distribution of the fit as a function of signal yield after it is convolved with a Gaussian distribution with the width of the systematic uncertainty.

The ratio $\Rpsi$ can be be determined from the above measurements.
By sampling the signal yields of $\X\to\gamma\jpsi$ and $\X\to\gamma\psip$ according to their likelihood distributions, a probability distribution that depends on $\Rpsi$ is obtained.
After convolving this with a Gaussian distribution representing the uncommon systematic uncertainty between the two channels, the UL on $\Rpsi$ is determined to be $0.59$ at the 90\% C.L.

We also perform fits where the signal contribution is fixed to the expectation calculated from previous measurements.
We fix the cross-section of $\EE\to\gamma\X,~\X\to\pp\jpsi$ production to the value reported in Ref.~\cite{X2pipiJpsi_BESIII} and take the relative ratio $\frac{\mathcal{B}(\X\to\gamma\psip)}{\mathcal{B}(\X\to\pp\jpsi)}$ from
a global fit~\cite{Li:2019kpj}, or fix $\X\to\gamma\jpsi$ to our own result and take $\Rpsi$ from an LHCb measurement~\cite{R_in_lhcb}, and  from a Belle measurement~\cite{R_in_belle}.
The results, also shown in Fig.~\ref{fig:FitX}(b),  have a goodness-of-fit of $\chi^2/{\rm ndf}=46.9/59~(p=0.87)$, $66.8/59~(p=0.23)$, and $46.0/59~(p=0.89)$ for the BESIII, LHCb and Belle hypotheses, respectively.
Our result for $\Rpsi$ is $2.8\sigma$ lower than that reported by the LHCb collaboration, corresponding to a $p$-value of 0.0048 calculated with $p=\int^{\infty}_{0}\int_{R_{0}}^{\infty}\mathcal{L}(R) G(R_{0}) dRdR_{0}$,
where $\mathcal{L}(R) $ is the likelihood distribution in this work and $G(R)$ is the Gaussian-assumed likelihood profile of the uncertainty of LHCb measurement.

\begin{figure}[htbp]
\raggedright
    \begin{overpic}[width=0.23\textwidth]{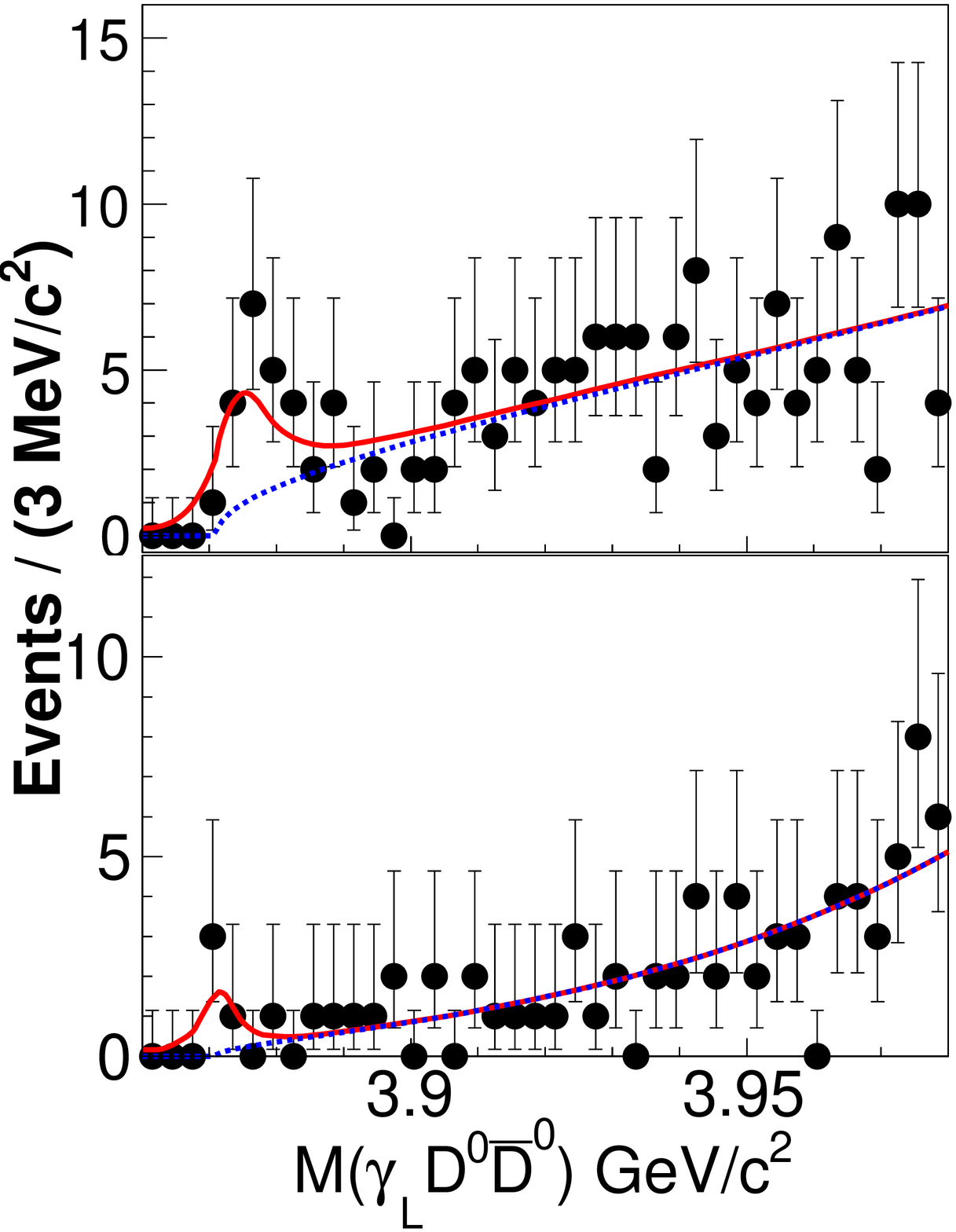}
        \put(25,80){\textbf{(a)}}
        \put(25,40){\textbf{(b)}}
    \end{overpic}
        \begin{overpic}[width=0.23\textwidth]{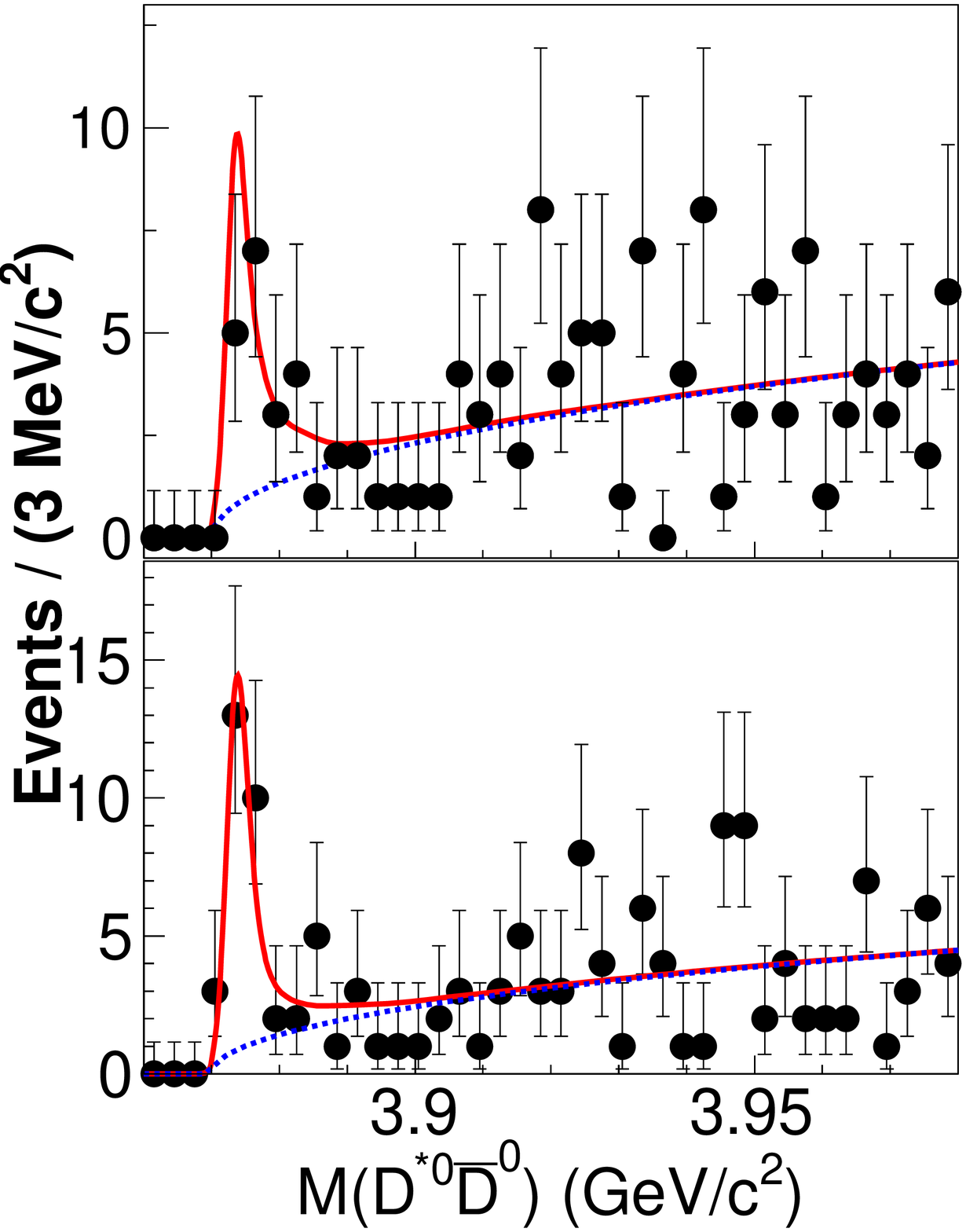}
        \put(25,80){\textbf{(d)}}
        \put(25,40){\textbf{(e)}}
    \end{overpic}
    \begin{overpic}[width=0.23\textwidth]{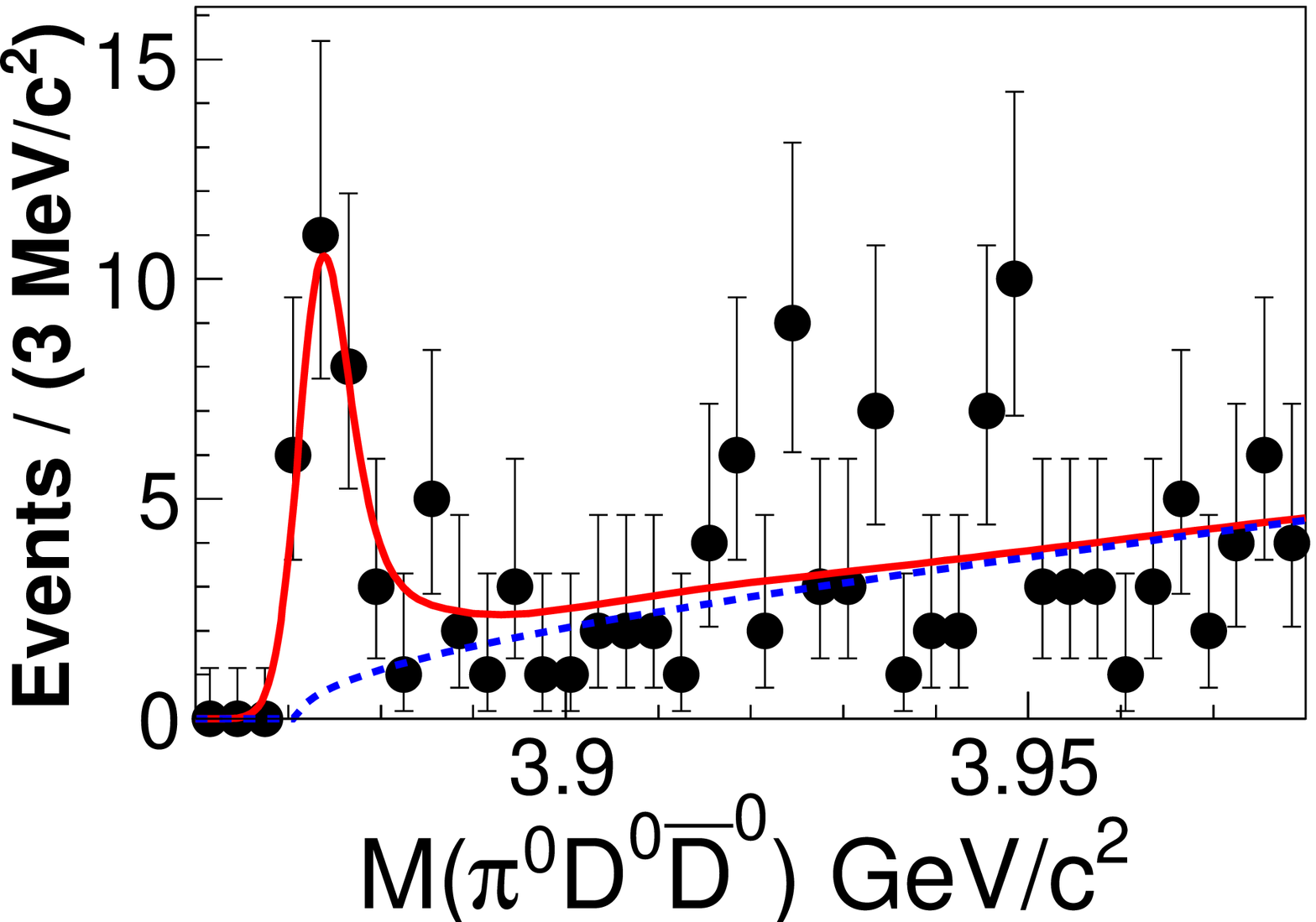}
        \put(30,50){\textbf{(c)}}
    \end{overpic}
    \begin{overpic}[width=0.23\textwidth]{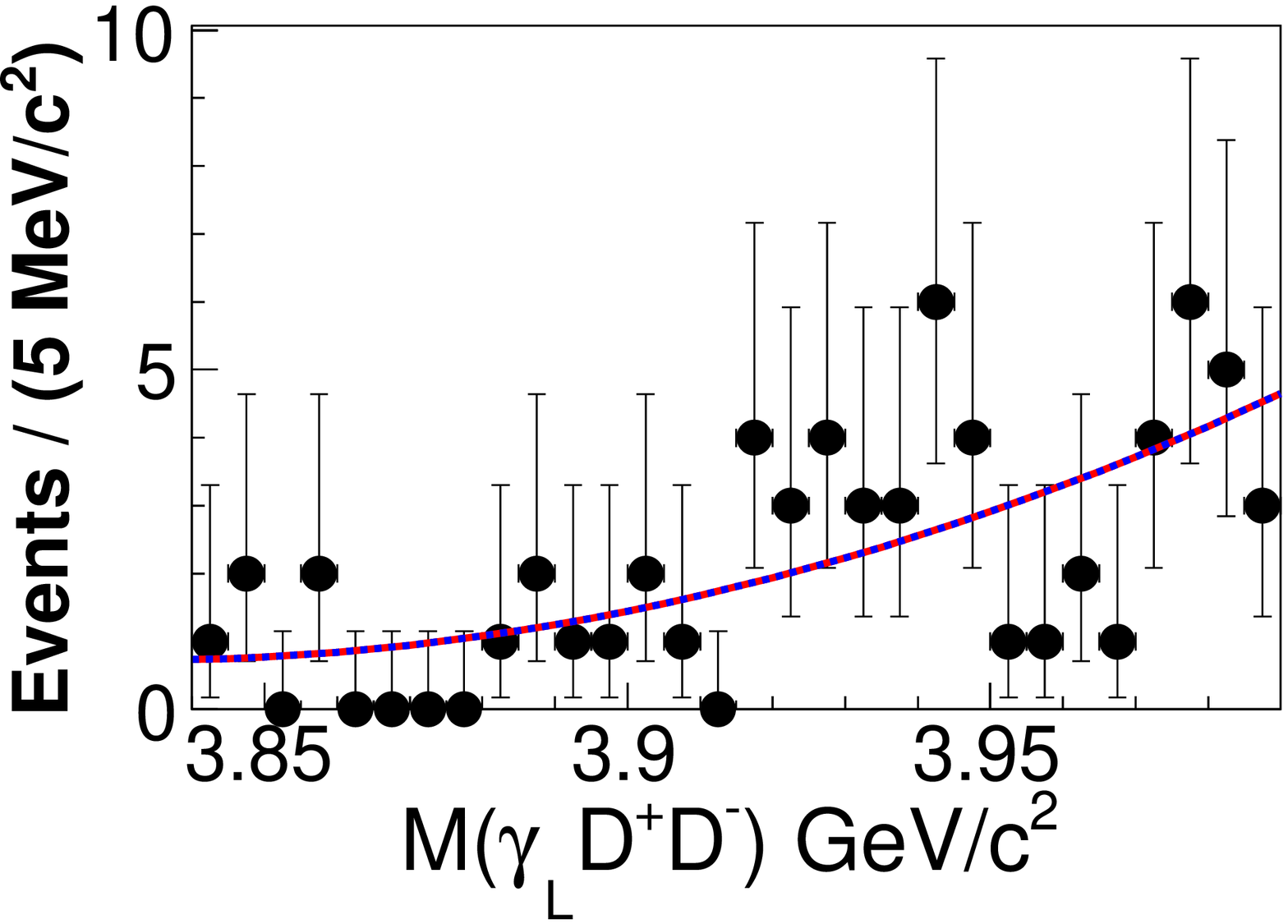}
        \put(30,50){\textbf{(f)}}
    \end{overpic}
    \caption{Left column: $M(\gammaL \DDbar)$ with $M(\gammaL D^{0})$ in (a) or below (b) the $D^{*0}$ mass window.
      $M(\pi^0\DDbar)$ with $M(\pi^0 D^0)$ in the $D^{*0}$ mass window (c).
      Right column: Simultaneous fit results for $\X\to\DDstar$ with $\Dstar\to\gamma D^0$ (d) and $\Dstar\to\pi^0 D^0$ mode (e).
      Fit results for $\X\to\gammaL\DpDm$ (f). The points with error bars are from data, the red curves are the best fit,
      and the blue dashed curves are the background components (color online). }
    \label{fig:FitX_DD}
\end{figure}

We consider the possibility of non-resonant three-body production to  the final states $\gamma\DDbar$ and $\pi^0\DDbar$, in addition to the well-established decay $\X \to \DDstar$.
We only search for $\X$ with $\gammaL\DDbar$ because the photon energy in $\X\to\gamma\DDbar$ is always lower than that in $\EE\to\gamma\X$.
The mass spectra $M(\gammaL\DDbar)$ and $M(\pi^0\DDbar)$ are shown in Fig.~\ref{fig:FitX_DD}
for the case when $M(\gammaL/\pi^0 D)$ lies in (a) or out of (b) the  $\Dstar$ mass region,
and when $M(\pi^0 \DDbar)$ lies in this mass range (c).
We fit the three mass spectra individually, and use an efficiency matrix determined from MC simulation that accounts for migrations of true events between the mass ranges, to determine the number of produced events in each category.
The signal yields for non-resonant three-body $\X\to\gamma\DDbar$ production and the decay $\X \to \DDstar~(D^{*0}\to\gamma D^0)$ are found to be $1.3\pm0.7$ and $20.5\pm7.4$, respectively, and
the corresponding yields  for  $\X\to\pi^0\DDbar$ and $\X \to \DDstar~(D^*\to\pi^0 D^0)$ decays are $-0.5\pm2.3$ and $36.1\pm7.7$, respectively.
The yields for the  three-body decays are not significant and so we set ULs at the 90\% C.L. of $8.7$ events for $\X\to\gamma\DDbar$ and $2.3$ events for $\X \to \pi^0\DDbar$, corresponding to $3.2\times10^2$ and $1.2\times10^2$ BF- and efficiency- corrected events, respectively.
Here systematic uncertainties, which are discussed later, are taken into account.

In the next stage of the analysis of the $\X\to\DDstar$ decays,
the combination of $\gammaL\DO$ or $\pi^0\DO$ with an invariant mass closest to the $D^{*0}$ nominal mass is taken as the $D^{*0}$ candidate.
For the channel $\Dstar\to\gamma\DO$,  the mass window for selecting the $\Dstar$ is
$M(\gammaL\DO)\in[m_{D^{*0}}-0.006,~m_{D^{*0}}+0.006]~\rm GeV/$$c^2$,
while for $\Dstar\to\pi^0\DO$ it is $M(\pi^0\DO)\in[m_{D^{*0}}-0.004,~m_{D^{*0}}+0.004]~\rm GeV/$$c^2$.
The distributions of the corrected invariant mass $M(\DDstar)\equiv M(\gamma(\pi^0)\DDbar)-M(\gamma(\pi^0)D)+m_{D^{*0}}$
are shown in Fig.~\ref{fig:FitX_DD}(d,e) following these requirements,
where contributions from non-resonant three-body processes are neglected.

To measure the $\X\to\DDstar$ signal, a simultaneous fit is performed to the corrected invariant-mass distributions.
The ratio of the signal yields for $\Dstar\to\gamma\DO$ and $\pi^0\DO$ is constrained to the product of corresponding BFs and averaged reconstruction efficiencies.
The signals are represented by MC simulated shapes, and the backgrounds by ARGUS functions~\cite{ARGUS}, with thresholds fixed at $m_{\Dstar}+m_{\bar{D^0}}$.
The fit results are shown in Fig.~\ref{fig:FitX_DD}(d,e).
The number of efficiency- and BF-corrected $\X\to\DDstar$ events is $(30.0\pm5.4)\times10^3$, and corresponds to
$20.2\pm3.6$ and $25.5\pm4.6$ observed events for $\Dstar\to\gamma\DO$ and $\pi^0\DO$ modes, respectively.
The goodness-of-fit is $\chi^2/{\rm ndf}=13.0/16~(p=0.67)$ after re-binning the data to satisfy the criterion that there are at least seven events in one bin.
Varying the fit range and describing the background with alternative shapes always results in a signal fit that has a statistical significance greater  than $7.4\sigma$.

The invariant mass of the $\gamma\DpDm$ system following the  $\X\to\gamma\DpDm$ selection
is shown in Fig.~\ref{fig:FitX_DD}(f).
No evident $\X$ signal is found.  This conclusion is quantified by performing an unbinned maximum-likelihood fit to the invariant-mass distribution, in which the signal component is described by a MC-simulated shape and the background is represented by a second-order polynomial.
The goodness-of-fit is $\chi^2/{\rm ndf}=6.2/5~(p=0.29)$.
The fit yields $(0.0^{+0.5}_{-0.0})$ $\X$ events. The UL on the number of the produced $\X\to\gamma\DpDm$ is $2.8\times10^3$ events at the 90\% C.L., with systematic uncertainties included in the calculation.

The decay channel $\X\to\pp\jpsi$  is reconstructed~\cite{X2pipiJpsi_BESIII,BESIII_omegaJpsi} to provide a normalization mode against which the rates of the other decays can be compared.
This channel yields $93.9\pm11.4$ $\X\to\pp\jpsi$ events, corresponding to $(24.9\pm3.0)\times10^{2}$ BF- and efficiency-corrected events.
The relative ratios can then be obtained by sampling the number of produced events of $\gamma\jpsi$,~$\gamma\psip$, $\gamma\DDbar$, $\pi^0\DDbar$, $\DDstar$, and $\gamma\DpDm$ according to the likelihood distributions, compared with that of $\pp\jpsi$.
We convolute the distributions with a Gaussian whose width is the systematic uncertainty of each channel, where uncertainties in common with the $\pp\jpsi$ channel are excluded.
The ratios are listed in Table~\ref{tab:ratio} for the modes studied in this paper, together with $\X \to \omega J/\psi$ and $\pi^0\chi_{c1}$, whose production rates have recently been measured by BESIII~\cite{BESIII_omegaJpsi,X2pi0chic1}.

\begin{table}[htbp]
\caption{Relative branching ratios and UL on branching ratios compared with $\X\to\pp\jpsi$~\cite{BESIII_omegaJpsi} \cite{X2pi0chic1}, where systematic uncertainties have been taken into account.} \label{tab:ratio}
\begin{tabular}{ c c c  }
\hline
\hline
\specialrule{0em}{1pt}{1pt}
    mode  & ratio & UL \\
    \hline
    $\gamma\jpsi$ 		&  $0.79\pm0.28$ 							&  -     \\
    $\gamma\psip$		&  $-0.03\pm0.22$	 						&  $<0.42$   \\
    $\gamma\DDbar$ 		&  $0.54\pm0.48$ 							&  $<1.58$   \\
    $\pi^0\DDbar$~    		&  $-0.13\pm0.47$ 	 						&  $<1.16$  \\
    $\DDstar+c.c.$~  		&  $11.77\pm3.09$	  						& - \\
    $\gamma\DpDm$ 		&  $0.00^{+0.48}_{-0.00}$  					&$<0.99$ \\
    $\omega\jpsi$  		&  $1.6^{+0.4}_{-0.3}\pm0.2$~\cite{BESIII_omegaJpsi}  &  - \\
    $\pi^0\chi_{c1}$ 		&  $0.88^{+0.33}_{-0.27}\pm0.10$~\cite{X2pi0chic1} & -  \\
\hline

\end{tabular}
\end{table}

Systematic uncertainties considered in the analysis include the detection efficiency, sub-decay BFs, mass window requirements, kinematic fit, ISR correction, generator model, and background shapes.
The uncertainties associated with the knowledge of the detection efficiency, including tracking efficiency (1\% per track), photon detection efficiency (1\% per photon), PID efficiency (1\% per track), $\pi^0$ reconstruction efficiency (1\% per $\pi^0$) are assigned following the results of earlier BESIII studies~\cite{sysTrk,sysPhn}.
The uncertainties listed for the modes that involve multiple sub-decays are calculated and weighted according to the BF and efficiency as well as the correlations between the different decay channels used to reconstruct these states.   The uncertainties on the
 BFs of the $D$ meson, $\jpsi$, and $\psip$ decays are taken from Ref.~\cite{PDG}.

The uncertainty associated with the mass window used to select $\jpsi$ mesons, which arises from a difference in resolution between data and MC, is 1.6\%~\cite{X2pipiJpsi_BESIII},
and that for selecting $D$ mesons is 0.7\% per $D$ meson~\cite{Y2piDDstar}.
The systematic uncertainty associated with the efficiency of the kinematic fit is estimated using the method discussed in Ref.~\cite{helix}.

To assign the systematic uncertainty associated with the MC events generation, we take the change in reconstruction efficiency when varying the assumption of an E1 transition in $\EE\to\gamma\X$ and $\X\to\gamma\jpsi(\psip)$ decays to pure phase space.
We change the energy-dependent cross-section lineshape of the $Y(4260)$~\cite{PDG} in the generator to the measured $\EE\to\gamma\X$~\cite{BESIII_omegaJpsi} lineshape and the difference on the reconstruction efficiency is taken as the systematic uncertainty due to the ISR correction.
To estimate the uncertainty arising from the limited knowledge of the  background shapes, we vary the shapes to different order of polynomials in the fit, and change the fit range at the same time.
To incorporate the systematic uncertainty into the UL, the most conservative result in the various fits is taken as the final result.
The effects on the modeling of the signal shapes from discrepancies between the mass resolution in data and MC simulation are negligible.

The systematic uncertainties of the kinematic fit (1\%), ISR correction (1\%), and background (4.0\%) in $\X\to\pp\jpsi$ mode are taken from Ref.~\cite{BESIII_omegaJpsi}.
A summary of the systematic uncertainties of the relative ratios is presented in supplemental material.
The common uncertainties have been cancelled, and the uncommon ones from $\X\to\pp\jpsi$ mode have been propagated into the results.
The total systematic uncertainty is obtained by adding the individual
components in quadrature.

In summary, using $\EE$ collision data taken at $\sqrt{s}=4.178$--$4.278$ GeV, we observe  $\X\to\DDstar+c.c.$ and find evidence
for $\X\to\gamma\jpsi$ with significances of $7.4\sigma$ and $3.5\sigma$, respectively.
No evidence is found for the decays $\X\to\gamma\psip$ and $\X\to\gamma\DpDm$.
The UL on the ratio $\Rpsi<0.59$ is obtained at the 90\% C.L.;
this is consistent with the Belle measurement~\cite{R_in_belle} and the global fit~\cite{Li:2019kpj}, but challenges the LHCb measurement~\cite{R_in_lhcb}.
Our measurement, taking into account model predictions, suggests that the X(3872) state is more likely a molecule or a mixture of molecule and charmonium, than a pure charmonium state.
We also measure the ratios of BFs for $\X\to\gamma\jpsi$, $\gamma\psip$, $\gamma\DDbar$, $\pi^0\DDbar$, $\DDstar+c.c.$, and $\gamma\DpDm$ to that for $\X\to\pp\jpsi$.
As discussed in Ref.~\cite{DDstarMolecule}, the relative ratios can be calculated on the assumption that the $\X$
is a bound state of $\DDstar$. We note, however, that no predictions are yet available for a binding energy of $(0.01\pm0.20)$~MeV,
which is the value that is obtained from the most recent mass measurements~\cite{PDG}.
Our measurement provides essential input to future tests of the molecular model for the $\X$ meson.

The BESIII collaboration thanks the staff of BEPCII and the IHEP computing center for their strong support. This work is supported in part by National Key Basic Research Program of China under Contract No. 2015CB856700; National Natural Science Foundation of China (NSFC) under Contracts Nos. 11905179, 11625523, 11635010, 11735014, 11822506, 11835012, 11961141012; the Chinese Academy of Sciences (CAS) Large-Scale Scientific Facility Program; Joint Large-Scale Scientific Facility Funds of the NSFC and CAS under Contracts Nos. U1532257, U1532258, U1732263, U1832207; CAS Key Research Program of Frontier Sciences under Contracts Nos. QYZDJ-SSW-SLH003, QYZDJ-SSW-SLH040; 100 Talents Program of CAS; INPAC and Shanghai Key Laboratory for Particle Physics and Cosmology; ERC under Contract No. 758462; German Research Foundation DFG under Contracts Nos. Collaborative Research Center CRC 1044, FOR 2359; Istituto Nazionale di Fisica Nucleare, Italy; Ministry of Development of Turkey under Contract No. DPT2006K-120470; National Science and Technology fund; STFC (United Kingdom); The Knut and Alice Wallenberg Foundation (Sweden) under Contract No. 2016.0157; The Royal Society, UK under Contracts Nos. DH140054, DH160214; The Swedish Research Council; U. S. Department of Energy under Contracts Nos. DE-FG02-05ER41374, DE-SC-0010118, DE-SC-0012069.


\begin{thebibliography}{99}


\bibitem{FirstObservationOfX}S.K. Choi {\it et al.} [Belle Collaboration], Phys. Rev. Lett. {\bf 91}, 262001 (2003).
\bibitem{JPC_X3872_1} A. Abulencia {\it et al.} [CDF Collaboration], Phys. Rev. Lett. {\bf 98}, 132002 (2007).
\bibitem{JPC_X3872_2} R. Aaij {\it et al.} [LHCb Collaboration], Phys. Rev. Lett. {\bf 110}, 222001 (2013).
\bibitem{DDstarMolecule} E.S. Swanson, Phys. Lett. B {\bf 598}, 197 (2004); E.S. Swanson, Phys. Rep. {\bf 429}, 243 (2006).
\bibitem{R_in_DDstar} J.~Ferretti, G.~Galata and E.~Santopinto, Phys.\ Rev.\ D {\bf 90}, 054010 (2014).
\bibitem{R_in_mix} E. J. Eichten, K. Lane, and C. Quigg, Phys. Rev. D {\bf 73}, 014014 (2006).
\bibitem{R_in_chm1} T. Barnes, S. Godfrey, and E.S. Swanson, Phys. Rev. D {\bf 72}, 054026 (2005).
\bibitem{R_in_chm2} T. Barnes and S. Godfrey, Phys. Rev. D {\bf 69}, 054008 (2004).
\bibitem{R_in_chm3} B.Q. Li and K.T. Chao, Phys. Rev. D {\bf 79}, 094004 (2009).
\bibitem{R_in_chm4} T. Lahde, Nucl. Phys. A {\bf714}, 183 (2003).
\bibitem{R_in_chm5} A.M. Badalian, V.D. Orlovsky, Y.A. Simonov, and B.L.G. Bakker, Phys. Rev. D {\bf 85}, 114002 (2012).
\bibitem{R_in_chm6} T. Mehen and R. Springer, Phys. Rev. D {\bf 83}, 094009 (2011).
\bibitem{R_in_chm7} T.M. Wang and G.L. Wang, Phys. Lett. B {\bf 697}, 3 (2011).
\bibitem{R_in_lhcb} R.~Aaij {\it et al.} [LHCb Collaboration], Nucl.\ Phys.\ B {\bf 886}, 665 (2014).
\bibitem{R_in_babar} B.~Aubert {\it et al.} [BaBar Collaboration], Phys.\ Rev.\ Lett.\  {\bf 102}, 132001 (2009).
\bibitem{R_in_belle} V.~Bhardwaj {\it et al.} [Belle Collaboration], Phys.\ Rev.\ Lett.\  {\bf 107}, 091803 (2011).
\bibitem{X2pipiJpsi_BESIII}M. Ablikim {\it et al}. [BESIII Collaboration], Phys.\ Rev.\ Lett.\ {\bf 112}, 092001 (2014).

\bibitem{BESIII_omegaJpsi}M.~Ablikim {\it et al.} [BESIII Collaboration],
  Phys.\ Rev.\ Lett.\  {\bf 122}, 232002 (2019).


\bibitem{Ablikim:2009aa}
  M.~Ablikim {\it et al.} [BESIII Collaboration],
  Nucl.\ Instrum.\ Meth.\ A {\bf 614}, 345 (2010).


\bibitem{etof}
 X.~Li {\it et al.}, Radiat. Detect. Technol. Methods {\bf 1}, 13 (2017);
 Y.X.~Guo {\it et al.}, Radiat. Detect. Technol. Methods {\bf 1}, 15 (2017).

\bibitem{geant4} S.~Agostinelli {\it et al.} [GEANT4 Collaboration], Nucl. Instrum. Meth. A {\bf506}, 250 (2003).

\bibitem{gmc}
  M.~Ablikim {\it et al.} [BESIII Collaboration],
  Phys.\ Rev.\ Lett.\  {\bf 123}, no. 11, 112001 (2019).



\bibitem{Zhang:2009bv}
  O.~Zhang, C.~Meng and H.Q.~Zheng,
  Phys.\ Lett.\ B {\bf 680}, 453 (2009).



\bibitem{PDG} M. Tanabashi {\it et al.} [Particle Data Group], Phys.\ Rev.\ D {\bf 98}, 030001 (2018).



\bibitem{ARGUS}
  H.~Albrecht {\it et al.} [ARGUS Collaboration],
  Phys.\ Lett.\ B {\bf 241}, 278 (1990).

\bibitem{Li:2019kpj}
  C.~Li and C.~Z.~Yuan,
  Phys.\ Rev.\ D {\bf 100}, no. 9, 094003 (2019).


\bibitem{sysTrk} M.~Ablikim {\it et al.} [BESIII Collaboration], Phys. Rev. D {\bf 93}, 011102 (2016)
\bibitem{sysPhn} M.~Ablikim {\it et al.} [BESIII Collaboration], Phys. Rev. D {\bf 87}, 012002 (2013).

\bibitem{Y2piDDstar}
  M.~Ablikim {\it et al.} [BESIII Collaboration],
  Phys.\ Rev.\ Lett.\  {\bf 122}, 102002 (2019).

\bibitem{helix}  M.~Ablikim {\it et al.} [BESIII Collaboration], Phys. Rev. D {\bf 87}, 012002 (2013).
\bibitem{X2pi0chic1}
  M.~Ablikim {\it et al.} [BESIII Collaboration],
  Phys.\ Rev.\ Lett.\  {\bf 122}, 202001 (2019).

\end{thebibliography}
\end{document}